\documentclass[a4paper,12pt,fleqn]{article}
\usepackage[dvipdfmx]{graphicx}
\usepackage{amsmath}
\usepackage{multirow}
\usepackage[latin1]{inputenc}             
\usepackage{bm}
\usepackage{setspace}
\usepackage{subfigure}
\usepackage{xcolor}
\usepackage{caption}
\usepackage{amssymb}
\usepackage{cite}
\usepackage{amsthm}
\usepackage{amsmath, amssymb}
\usepackage{type1cm}

\begin{document}

\title{Conditions for Robustness and Limitation on Bayesian Student-$t$ Linear Regression Modeling
}

\author{Yoshiko Hayashi \footnote{yoshiko-hayashi@omu.ac.jp
}\\
 Osaka Central Advanced Mathematical Institute\\
 Osaka Metropolitan University }

\maketitle

\begin{abstract}

The Student-$t$ linear regression model is a common method for robust modeling for addressing the outlier problem in linear regression models. It has been widely employed in several studies. However, before the model is applied, outliers and non-outliers need to be distinguished. This study provides practically useful and simple conditions to distinguish outliers from non-outliers. Thereafter, we establish sufficient conditions to ensure that the Student-$t$ linear regression model is partially robust against multiple outliers in the $y$-direction. \\
 
\it{Keywords\rm: Outlier, Linear regression model, Regularly varying distributions, Bayesian modeling}
\end{abstract}

\rm

\newpage
\section{Introduction}
\label{intro}

In regression analysis, outliers in a linear regression model can affect the accuracy of results obtained using the ordinary least squares (OLS) estimator. The Student-$t$ linear regression model, designed as a linear regression model with the error term having a \emph{t}-distribution, is a common method to solve the outlier problem (Lange et al., 1989). Although this model has been widely adopted, most studies apply it without careful theoretical consideration. 
 
 Bayesian robustness modeling using heavy-tailed distributions, which include a $t$-distribution, provides a theoretical solution to the outlier problem. 
  For a simple Bayesian model, when both the
prior distribution and likelihood of an observation are normal distributions, the
posterior distribution is also a normal distribution and the posterior mean
is a weighted average of the mean of the prior distribution and observation.
When the prior distribution and observation are further from each other and follow a normal distribution, the posterior distribution is far from both pieces of information; this is known as $conflict$. For instance, when a single observation $x$=15 follows $N(\mu ,1)$ and the prior of the location parameter follows $N(0,1)$, the posterior distribution follows $N(7.5, 0.5)$.
In this case, the posterior distribution is not determined by
either the prior distribution or observation.

 To address this issue, Dawid (1973) formally provided the theoretical resolution of the conflict between the prior distribution and data, also known as \it{conflict of information}. \rm{The author used a pure location model in which the scale parameter was provided and clarified how an outlier can be automatically ignored in the posterior distribution when the outlier follows a heavy-tailed distribution. This result occurs because information about a prior distribution is more credible than the observation.
 
O'Hagan (1990) proposed the concept of $credence$, which measures the
degree of the tail's information. Andrade and O'Hagan (2006) defined credence as the extent to which a source of information is more credible than another in a conflict; this is represented as the index of a regularly varying function. 
 Andrade and O'Hagan (2011) showed that in a univariate model, multiple observations located sufficiently close create a larger credence, which equals the sum of each credence of the observations. When an outlier is far from the group of non-outliers with the same heavy-tailed distribution, the information of the group of non-outliers creates larger credibility or credence. Thus, the posterior distribution is located closer to the non-outliers and is robust against the outlier.
  Andrade and O'Hagan (2011) established sufficient conditions for robust modeling against a single outlier in $n$ samples for a univariate model using the regular variation theory. The sufficient condition requires the minimum number of non-outliers to be robust against an outlier. O'Hagan and Pericchi (2012) reviewed previous studies on the resolution of the conflict. 

O'Hagan (1988) applied heavy-tailed modeling to a Student-$t$ linear regression model without an intercept term under the pure location structure and demonstrated its robustness. For a model without an intercept term, the outlier unconditionally conflicts with non-outliers. Therefore, a univariate model can be directly applied. By contrast, as Pe\~na et al. (2009) mentioned, the outlier in the $x$-direction for a model with an intercept term should be carefully addressed. Pe\~na et al. (2009) showed that when the outliers in the $x$-direction reached infinity, the result of a Student-$t$ linear model did not enable robustness. Pe\~na et al. (2009) examined the phenomenon using Kullback--Leibler divergence and proposed a down-weighting method that assigned a lower weight to outliers. Andrade and O'Hagan's (2011) showed that heavy-tail modeling using $t$-distribution is a partial robust modeling. Thus, a location-scale model cannot completely ignore outliers. Gagnon et al. (2020) theoretically developed a robust linear regression model using a super heavy-tailed distribution, which is heavier than the $t$-distribution and provides wholly robust modeling. He et al. (2021), Andrade (2022), and Gagnon and Hayashi (2023) provided theoretical considerations for the Student-$t$ linear regression model. Although the Student-$t$ linear regression model provides partial robustness, it is widely applied. Thus, clarification regarding the working mechanism of the model as a robust model is necessary.  

This study investigated the conditions for the Student-$t$ linear model with an intercept term for an outlier in the $y$-direction by extending Andrade and O'Hagan's (2011) conditions. First, we investigated the range in which conflict exists between an outlier and non-outliers, as the necessary condition to apply heavy-tail modeling. Then, we clarified the condition of the model's robustness.
    \begin{figure}[h]
\begin{center}
\includegraphics[ width=75mm]{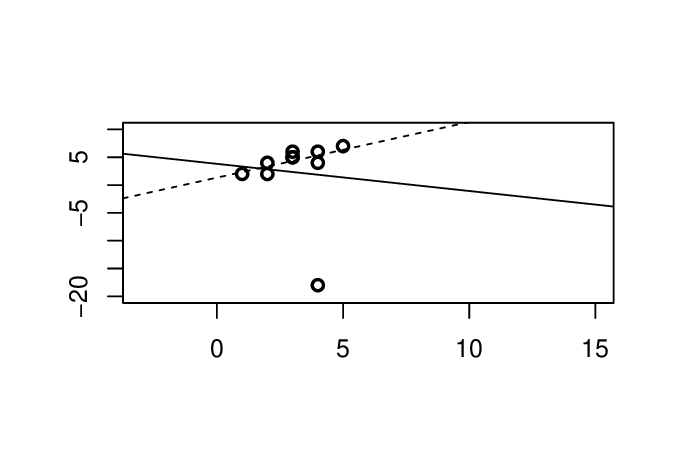}
\includegraphics[ width=75mm]{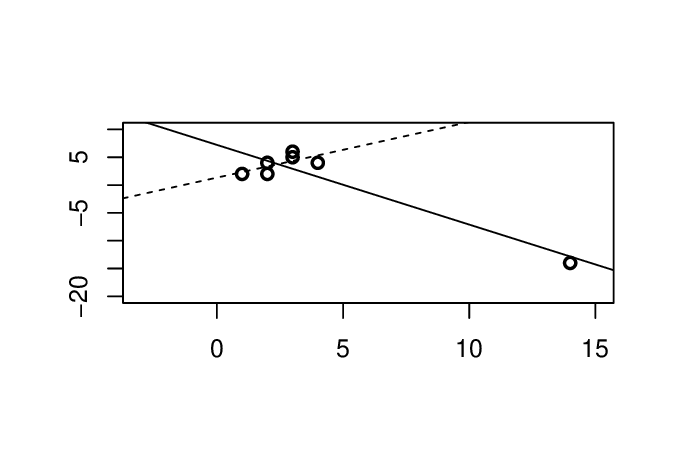}
\vspace{0cm}
\caption{Conflict in linear model: Straight lines show the regression line by OLS and dotted lines show it without the outlier.}
\end{center}
\end{figure}
Heavy-tailed modeling, as a resolution of conflict between an outlier and non-outliers, is effective when the outlier and the mean of the group of non-outliers are located sufficiently far and the sufficient condition for the number of non-outliers is satisfied. A linear regression model provides the mean of $y$ conditioned on $x$. Thus, the conflict of information in a linear regression model with an intercept term occurs when an outlier is located far from the regression line and non-outliers lie close to the regression line created from the non-outliers. 

The left panel in Figure 1 shows the case in which the outlier conflicts with the group of non-outliers. The figure shows that the outlier is located far from the regression line obtained using OLS. In this case, the Student-$t$ linear regression model is robust against the outlier in the $y$-direction. This is because the information of the conditional distribution of the outlier is less credible than that of the grouped non-outlier data, under the assumption of the same degrees of freedom (DoFs) of the $t$-distribution for all data, which represent $credence$. Non-outliers in the left panel of Figure 1 are close to each other and create a large credence, whereas the outlier does not belong to the regression line generated by the grouped data and creates a small credence.
Meanwhile, as shown in the right panel of Figure 1, when the outlier is in the $x$-direction, which is known as the leverage point, all data, including the outlier, are sufficiently close to the regression line and create a larger credence than that of the regression line without the outlier, as presented by the dotted line in Figure 1. In this case, the straight line in the right panel of Figure 1 has a larger credence than that of the dotted line.

The remainder of this paper is organized as follows: Section 2  presents the condition for the existence of conflict between an outlier and non-outliers in the Student-$t$ linear regression model. Section 3 highlights the sufficient conditions for the Student-$t$ linear regression model. Section 4 presents simulation results in a simple linear regression model, and Section 5 presents the conclusions.

\section{Conflicting Information in the Student-$t$ Linear Regression Model}

To examine the limitation of the robustness of the Student-$t$ linear model with an intercept term, the following linear regression model was considered. The dependent variable $ \bf y$ is an $n \times 1$ vector, the independent variable $\bf X$ is an $n \times (k+1)$ full-rank matrix and fixed, $\bf \beta$ is a $(k+1) \times 1$ vector, and $\bf u$ is an $n \times 1$ vector assumed to be independent and identically distributed random errors: 
 \begin{eqnarray}
\bf{ y =X \beta + u}, 
\end{eqnarray}
where

\[
\bf  X_{} \it = 
  \left[
    \begin{array}{cccc}
     1 & X_{11} &\ldots & X_{d1} \\
      \vdots & \vdots & \ddots & \vdots \\
      1 & X_{1n} & \ldots & X_{dn}
    \end{array}
    \right].\\
    \]

Consider the residual of the result from OLS for the model in Equation (1):
 \begin{eqnarray}
\bf{ y =X \hat{\beta}_{ols} + e},
\end{eqnarray}
where
\[
\bf  e'_{} \it = 
  \left[
    \begin{array}{c}
    e^{1}_{/out}, \ldots , e^{n-1}_{/out},e_{out}
    \end{array}
    \right],\\
    \]\\
    and subscripts ${/out}$ and ${out}$ denote a non-outlier and an outlier, respectively.
   
According to Cook and Weiberg (1982) and Chatterjee and Hadi (1988), the prediction or hat matrix, $\bf \hat{y}=Hy$, is given by
 \begin{eqnarray}
 \label{Eq2.2}
\bf H=X(X^{\it{T}}\bf X)^{\rm{-1}}\bf X.
\end{eqnarray}
The residual is given as
\begin{eqnarray}
\label{Eq2.3}
\bf{e}=(\bf{I}_{\it n}-\bf{H})\bf{y}.
\end{eqnarray}
For a model with an intercept term, the ($i,j$)-th element of $\bf H$, the elements $h_{ij}$ of the Hat matrix is given as
 \begin{eqnarray}
 \label{Eq2.4}
 h_{ij}=\frac{1}{n}+ \bf (x_{\it{i}}-\bar{x})^{\it{T}}\bf(\tilde{X}^{\it{T}}\tilde{X})^{-1}(x{_{\it{j}}}-\bar{x})
\quad (\it i,j=\rm 1,2,\ldots, \it n),
  \end{eqnarray}
 where 
\[
  \bf \widetilde{X_{}} \it = 
  \left[
    \begin{array}{ccc}
     X_{11}-\bar{X}_{1} &\ldots & X_{d1}-\bar{X}_{d} \\
      \vdots & \ddots & \vdots \\
      X_{1n}-\bar{X}_{1} & \ldots & X_{dn}-\bar{X}_{d}
    \end{array}
  \right],\\
  \bf x_{\it{i}} = 
  \left[
    \begin{array}{c}
     X_{1i} \\
      \vdots \\
      X_{di}
    \end{array}
  \right],\\
    \bf \bar{x} \it = 
  \left[
    \begin{array}{c}
     \bar{X}_{1}\\
      \vdots \\
      \bar{X}_{d} 
    \end{array}
  \right],\\
  \]  
and $\bar{X}_{m}=\Sigma_{i=1}^{n}X_{mi}/n \quad (m=1, \ldots, d)$.

  For a model with an intercept term, for the diagonal elements, which are known as $leverage$, $1/n$ is the smallest value
  and 1 is the largest.
  \begin{equation}
  \label{Eq2.5}
\frac{1}{n} \leq h_{ii} \leq 1 \quad (\it i=\rm 1,2,\ldots, \it n).
\end{equation}
 Mohammadi (2016) expressed the range of the off-diagonal elements with an intercept term as follows:
 \begin{equation}
 \label{Eq2.6}
\frac{1}{n}-\frac{1}{2} \leq h_{ij} \leq \frac{1}{2} \quad (\it i \neq j).
\end{equation}
Assume that the non-outliers are located sufficiently close to the regression line.
If the outlier moves away from the regression line faster than the group of non-outliers, the residual $e_{out}$ reaches infinity as the outlier reaches infinity in the $y$-direction. Because non-outliers create combined credence, if one of the non-outliers conflicts with the outlier, the group of non-outliers conflicts with the outlier. As shown in Figure 1, when an outlier is located sufficiently close to the group of non-outliers in the $x$-direction, they conflict. Therefore, if the partial derivative of $e_{out}$ with respect to $y_{out}$ is greater than that of the residual of the closest non-outlier, the outlier conflicts with the group of non-outliers.

The relationship between the Hat matrix and range of outliers in a linear regression model can be derived as follows: Let the $n$-th observation be an outlier, $h^{out}_{nn}$ represents the element of $\bf H$ of an outlier.

\begin{eqnarray}
\frac{d e_{out}}{d y_{out}}>\frac{d e^{max}_{/out}}{d y_{out}},
\end{eqnarray}
where the subscript ${/out}$ is a non-outlier.

\rm From the definition, the derivatives of the residuals of an outlier, $e_{out}$, and a non-outlier, $e^{j}_{/out}$, with respect to an outlier, $y_{out}$, can be given as
\begin{eqnarray}
 \label{Eq2.7}
\frac{d e_{out}}{d y_{out}}=1-h^{out}_{nn}, 
\end{eqnarray}
and
\begin{eqnarray}
\label{Eq2.8}
\frac{d e^{j}_{/out}}{d y_{out}}=-h_{nj}.
\end{eqnarray}
To investigate the limitation in which an outlier conflicts with a group of non-outliers, this study examined the location of an outlier as it approaches infinity in the $y$-direction. To examine the condition, we first check whether the observation is an outlier in the first stage and subsequently derive the condition if the non-outliers create a group.\\

\noindent
\bf{Lemma 1}
\label{Lem1}
\it
If the following condition holds, the observation is an outlier in the linear regression model as $y_{out} \rightarrow \pm \infty$ :
\begin{eqnarray}
\label{Eq2.9}
h_{nj}> \rm 0 \quad  \it( j=\rm{1},\ldots, \it n-\rm 1). 
\end{eqnarray}

\noindent
Proof.
\rm
The condition that the observation is an outlier is expressed as the outlier moves away from the regression line, where the residual $e_{out}$ approaches infinity as the outlier turns to infinity in the $y$-direction. 
\begin{eqnarray}
 \label{Eq2.10}
\it \frac{d e_{out}}{d y_{out}}>\frac{d e^{j}_{/out}}{d y_{out}} \quad\quad &\quad \quad   (j=1,\ldots,n-1). 
\end{eqnarray}
 Substituting Equations (9) and (10) into condition (12), we have
\begin{eqnarray}
 \label{Eq2.11}
1-\it h^{out}_{nn}>-h_{nj}. 
\end{eqnarray}
Using (6), Lemma 1 is obtained. \\

\qed\\

\noindent
\bf{Lemma 2}
\label{Lem2}
\it
If the following condition holds, the non-outliers create a group against the outlier as $y_{out}$ approaches infinity in the linear regression model.
\begin{eqnarray}
\label{Eq2.12}
h^{out}_{nn}<\frac{1}{2}.
\end{eqnarray}
\rm

\noindent
\it Proof. \rm
The outlier moves away from the regression line faster than the group of non-outliers. 
 Therefore, if the derivative of $e_{out}$ with respect to $y_{out}$ is greater than that of non-outlier residuals, the non-outliers create a group against the outlier. Creating a group of non-outliers requires the following conditions:
\begin{eqnarray}
 \label{Eq2.13}
\begin{cases}
\rm (i) \quad \it \frac{d e_{out}}{d y_{out}}>\frac{d e_{/out}}{d y_{out}} \quad\quad &\rm  for \quad \it \frac{d e_{/out}}{d y_{out}}\geq\rm0,\\
\\
\rm (ii) \quad \it \frac{d e_{out}}{d y_{out}}>-\frac{d e_{/out}}{d y_{out}} \quad\quad&\rm for \quad \it \frac{d e_{/out}}{d y_{out}}<\rm0.
\end{cases}
\end{eqnarray}
When Lemma 1 holds, condition (i) in (15) holds: To satisfy condition (ii) in (15), we have 
\begin{eqnarray}
1-\it h^{out}_{nn}>h_{nj}
\end{eqnarray}
 for $d e_{/out}/d y_{out}<\rm0$. Lemma 2 is satisfied for the range of $h_{nj}$. 

\qed\\

\noindent
\bf{Corollary 1}
\label{Coro1}
\it
When the following condition holds, the property of idempotent and symmetric matrix is satisfied under the conditions of Lemmas 1 and 2.
\begin{eqnarray}
\Sigma_{d \neq j}^{n-1}(h_{nd})^2 \geq \frac{1}{8}   \quad \quad (j=1, \ldots, n-1).
\end{eqnarray}
\it

\noindent
Proof.  
\rm
According to Chatterjee and Hadi (1988), because a Hat matrix is idempotent and symmetric, $h^{out}_{nn}$ can be expressed as
\begin{eqnarray}
\label{Eq2.12}
h^{out}_{nn}&=&(h^{out}_{nn})^2+\Sigma_{j=1}^{n-1}(h_{nj})^2  \nonumber \\
               &=&(h^{out}_{nn})^2+(h_{nj})^2+\Sigma_{d \neq j}^{n-1}(h_{nd})^2 
\end{eqnarray}
By arranging (18), we have
\begin{eqnarray}
\label{Eq2.12}
\frac{1}{4}&=&\left(h^{out}_{nn}-\frac{1}{2}\right)^2+(h_{nj})^2+\Sigma_{d \neq j}^{n-1}(h_{nd})^2.
\end{eqnarray}
The dashed and dot-dashed lines in Figure 2 represents Lemmas 1 and 2 based on conditions (13) and (16). The dotted circle in Fig. 2 represents the case with $\Sigma_{k \neq j}^{n-1}(h_{nk})^2 =0$, which does not satisfy both Lemmas 1 and 2 for $h_{nn}^{out}>0.5$. From the figure, $\Sigma_{k \neq j}^{n-1}(h_{nk})^2$ is necessarily greater than or equal to $\frac{1}{8}$ to satisfy the conditions of Lemmas 1 and 2.

\qed

\begin{figure}[h]
\centering
\includegraphics[ width=100mm]{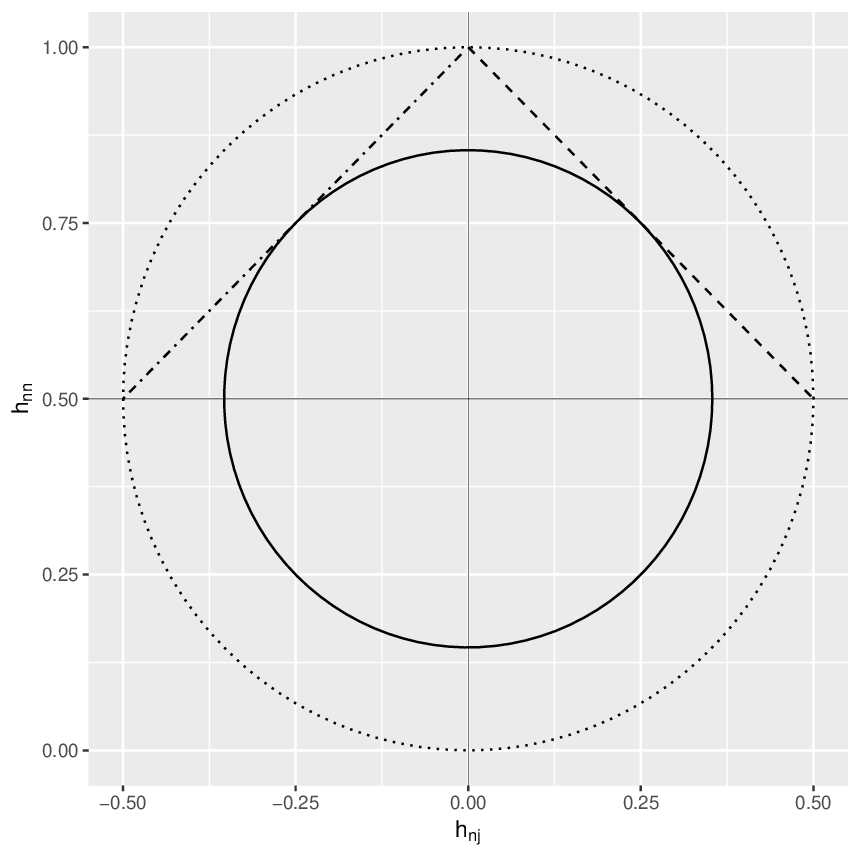}
\caption{Dashed and dot-dashed lines represent the conditions of Lemmas 1 and 2, respectively. The circle drawn by a line shows the case with $\Sigma_{d \neq j}^{n-1}(h_{nd})^2= \frac{1}{8}$, whereas that drawn by a dotted line shows $\Sigma_{d \neq j}^{n-1}(h_{nd})^2= 0$.}
 \label{Fig1}
\end{figure}
\clearpage

\section{Sufficient Conditions for Rejecting Outliers in the Student-$t$
Linear Regression Model}

\rm{This section investigates the sufficient conditions for a Student-$t$ linear regression model to be robust based on Andrade and O'Hagan's (2011) corollary 4, which elucidates the conditions for robustness against a single outlier out of $n$ samples in a univariate model. To examine these conditions, we adopt the independent Jeffreys priors derived by Fonseca et al. (2008, 2014) under the given DoFs.

As discussed by Andrade and O'Hagan (2006), the credence is defined as $c$ for $f(x)\in R_{-c} (c>0)$, where $R_{-c}$ denotes that $f(x)$ is regularly varying at $\infty$ with index $c$. The credence of a $t$ distribution with $\gamma$ DoFs is $\gamma+$1 (see Appendix A).

In this section, we assume that we have access to a data set of the form $(X_i, y_i)_{i = 1}^n$, where $X_i := (x_{i1}, \ldots, x_{ip})^T\in \mathbb{R}^p$ are $n$ vectors with data points from $p$ covariates and $y_i \in \mathbb{R}$ are $n$ observations of a dependent variable, with $n$ and $p$ being positive integers. Herein, the dependent variable is modeled using the covariates, and a Bayesian linear regression model is assumed. We consider $x_{i1}  = 1$ as an intercept in the model and $X_i$ as fixed and known vectors. 

In a linear regression, the random variable $y_i$ is modeled as $y_i = X_i^T \beta + u_i$ , $i = 1,\ldots, n$, where $\beta := (\beta_1, \ldots, \beta_p)^T \in \mathbb{R}^p$ is the vector of the regression coefficients, $\sigma > 0$ is a scale parameter, and $u_1, \ldots, u_n$ are random errors. We assume that $\beta$ and $\sigma$ are independent. 

Assume that all observations are $t$-distributed with DoFs, $\gamma$, $t_{\gamma}(\mu, \sigma)$ has mean $\mu$ and scale parameter $\sigma$ with $\gamma$ DoFs. As the $t$-distribution is a location-scale family, the likelihood can be denoted as $f(y_i|\bf X,\beta, \it \sigma)= \rm 1/ \it\sigma \times h [ (y_i- X'_i \beta)/\sigma ] $. $X'_i$ denotes the $i$-th row of $\bf X$. 
For simplicity, we assume that all observations have the same likelihood function and the non-outliers are sufficiently close to the conditional mean $X'_i\beta$.

We start from an outlier model with $(n-1)$ non-outliers, where $n \geq 3$ and $p \geq 1$.
Let  $y_{i} = a_{i} + b_{i} \omega$,for $i = 1, \dots, n$, where $a_{i}, b_{i} \in \mathbb{R}$ are constants, and $\omega \rightarrow \infty$. $b_{i} = 0$ if the data point is a non-outlier and $b_{i} \neq 0$ if it is an outlier. $\pi(\beta)$ and $\pi(\sigma)$ represent our priors for $\beta$ and $\sigma$, respectively. \\

 \noindent
 \bf{Model} \rm{
 \begin{equation}
\label{Eq333}
\left\{
\begin{array}{@{\,}ll}
y_i| \textbf{X},\beta, \sigma \stackrel{D}{\sim} t_\gamma(y_i|X_i,\beta,\sigma)=1/\sigma \times h [ (y_i-X^T_i\beta)/\sigma ]\hspace{0.3cm}\textrm{independent}\hspace{0.3cm}(i=1,\ldots,n),\\
\beta_q \stackrel{D}{\sim} \pi(\beta_q)\propto 1, \hspace{0.8cm} (q=1,\ldots,p),\\
\sigma  \stackrel{D}{\sim}\pi(\sigma) \propto 1/\sigma, \\
h \in R_{-(\gamma+1)}, \gamma>0, \hspace{0.8cm} (i=1,\ldots,n).
\end{array}
\right. 
\end{equation}

According to Andrade and O'Hagan (2011), the limiting posterior joint distribution is expressed as follows:
\begin{equation}
\label{Eq334}
 \lim_{\omega \rightarrow \infty} \pi(\beta, \sigma|  \textbf{X},\textbf{y})=\lim_{\omega \rightarrow \infty} \pi(\sigma| \textbf{X},\textbf{y})\cdot \lim_{\omega \rightarrow \infty} \pi(\beta| \sigma, \textbf{X},\textbf{y}).
\end{equation}
We treat $\textbf{X}$ as fixed and known, but include it to clarify the condition on $\beta$.
We need to derive the limits. The first term in the RHS relates to rejecting the scale parameter, and the second term refers to rejecting the location parameter given the scale parameter.\\

}
\color{black}

\noindent  
\bf{Theorem 1 (Robustness of an outlier among $n$ observations).}
\it  Consider data \rm$\textbf{y}=(\it y_1, \ldots,y_n)$ following model \rm (\ref{Eq333})\it, \color{black} and that the $n$-th observation is an outlier $(b_n \neq 0 )$ and the other observations are non-outliers $(b_n = 0 )$. If Lemmas 1 and 2, Corollary 1, and the following conditions hold:

\rm(1a)  $(\gamma+1) < \{  (n-1) \cdot (\gamma+1) \}$,

\rm(1b)   $\gamma <  (n-p-1)$;   

\color{black}
then, the posterior distribution partially ignores the outlier:
\begin{equation}
\label{Eq3.4}
\pi(\beta,\sigma|\mathbf{X}, \mathbf{y}) \propto \sigma^{\gamma}\pi(\beta, \sigma | \bf {X^{\rm (\it n- \rm 1)}}, \mathbf{y}^{\rm (\it n- \rm 1)}) \hspace{1cm} \textrm{as} \hspace{0.3cm} \it \omega \rightarrow \infty,
\end{equation}
 
 \it
\noindent
Proof. 
\rm
According to Andrade and O'Hagan (2011), the posterior distribution can be described as
\begin{eqnarray}
\label{Eq3.5}
\pi(\beta,\sigma|\mathbf{X}, \mathbf{y}) \propto\frac{1}{\sigma}h[(y_i-X^T_i\beta)/\sigma] \cdot \pi(\beta|\sigma, \textbf{X}^{(n-1)},\textbf{y}^{(n-1)}) \cdot \pi(\sigma| \textbf{X}^{(n-1)},\textbf{y}^{(n-1)}).
\end{eqnarray}

When scale parameter $\sigma$ is given, the posterior distribution of $\beta$ in the model is as follows:
\begin{eqnarray}
\label{Eq3.5}
\pi(\beta|\sigma, \textbf{X}^{(n-1)},\textbf{y}^{(n-1)}) &\propto& \pi(\beta) \cdot \Pi_{i=1}^{n-1}h[(y_i-X^T_i\beta)/\sigma]\\ \nonumber
&\propto&  \Pi_{i=1}^{n-1}h[(y_i-X^T_i\beta)/\sigma] \in R_{- (n-1) (\gamma+1)}.
\end{eqnarray}

Thus, as $h$ (the likelihood of the outlier) $\in R_{- (\gamma+1)}$, condition (1a) is always satisfied. By applying Theorem 2 in Andrade and O'Hagan (2011), the likelihood of the outlier is rejected by the likelihood of non-outliers, which is  shown in   (\ref{Eq3.5}) serves as the role of the prior in the Theorem.\color{black}

Applying the transformation $\tau=\frac{1}{\sigma}\beta$, which is a $p \times$ 1 vector, yields
\begin{eqnarray}
\label{Eq3.6}
\pi(\textbf{y}^{(n-1)}|\sigma, \textbf{X}^{(n-1)})&=& \left(\frac{1}{\sigma}\right)^{n-p-1}   \int_{\mathbb{R}^{p}} \Pi_{i=1}^{n-1} h[(y_i/\sigma)-X^T_i \tau]  d\tau.
\end{eqnarray}
When all elements of $\textbf X$ are given and bounded, $ \int_{\mathbb{R}^{p}}    \Pi_{i=1}^{n-1}h[(y_i/\sigma)-X^T_i \tau]  d\tau$ in $\sigma$ is $O$(1). Thus, as a function of $\sigma$, it is slowly varying,
\begin{eqnarray}
\label{Eq3.7}
 \int_{\mathbb{R}^{p}}   \Pi_{i=1}^{n-1}h[(y_i/\sigma)-X^T_i \tau]  d\tau  \in R_{0}.
\end{eqnarray}
Thus, the marginal posterior distribution of $\sigma$ given information $\textbf{X}^{(n-1)}$ and $\textbf{y}^{(n-1)}$ becomes
\begin{eqnarray}
\label{Eq3.8}
\pi(\sigma| \textbf{X}^{(n-1)},\textbf{y}^{(n-1)})&\propto&\pi(\sigma) \cdot \pi(\textbf{y}^{(n-1)}|\sigma, \textbf{X}^{(n-1)})\in R_{- (n-p)}. 
 \end{eqnarray}
Again, applying the transformation $\tau=\frac{1}{\sigma}\beta$ yields the marginal posterior distribution of $y_n$ given information $\textbf{X}^{(n-1)}$ and $\textbf{y}^{(n-1)}$ as 
\begin{eqnarray}
\label{Eq3.9}
f(y_{n}|\sigma, \textbf{X}^{(n-1)},\textbf{y}^{(n-1)})  \hspace{8cm}\\
\hspace{2.5cm}= \left(\frac{1}{\sigma}\right)^{n-p} \int_{\mathbb{R}^{p}}  h[(y_n/\sigma)-X^T_n \tau] \Pi_{i=1}^{n-1}h[(y_i/\sigma)-X'_i \tau]  d\tau. \nonumber 
\end{eqnarray}

When non-outliers are located sufficiently close to the regression line, Andrade and O'Hagan's (2011) Proposition 1, which gives the convolution of regularly varying densities being distributed as their sum, $f*g(x) \sim f(x)+g(x)$, can be applied
as $f(y)=h[(y_n/\sigma)-X^T_n \tau] $ and $g(y)=\Pi_{i=1}^{n-1}h[(y_i/\sigma)-X^T_i \tau]$. Because $min((\gamma+1),(n-1)(\gamma+1)) =(\gamma+1)$ for $n \geq 3$, when the residual $e_n=y_n-X^T_n\beta$ reaches infinity as $y_n$ turns to infinity, we obtain
 \begin{eqnarray}
 \label{Eq3.10}
\int_{\mathbb{R}^{p}}  h[(y_n/\sigma)-X^T_n \tau] \Pi_{i=1}^{n-1}h[(y_i/\sigma)-X^T_i \tau]  d\tau\in R_{- (\gamma+1)}.  
\end{eqnarray}
Lemmas 1 and 2 and Corollary 1 provide the conditions for the residual $e_n=y_n-X^T_n\beta$ reaching infinity as $y_n$ approaches infinity.
Accordingly, the marginal posterior distribution for $\sigma$ is 
 \begin{eqnarray}
 \label{Eq3.11}
 \pi(\sigma| \textbf{X}, \textbf{y})=\frac{f(y_{n}|\sigma, \textbf{X}^{(n-1)},\textbf{y}^{(n-1)})\pi(\sigma|X^{(n-1)},\textbf{y}^{(n-1)})}{\int_{0}^{\infty} f(y_{n}|\sigma,  \textbf{X}^{(n-1)},\textbf{y}^{(n-1)})\pi(\sigma| \textbf{X}^{(n-1)},\textbf{y}^{(n-1)}) d\sigma}.
\end{eqnarray}
Next, consider the case where $y_n$ or $\omega$ approaches infinity.
As a function of $y_n$, the posterior distribution of $f(y_{n}|\sigma, \textbf{X}^{(n-1)},\textbf{y}^{(n-1)})$ assumes the form $\frac{1}{\sigma}g\left(\frac{y_n}{\sigma}\right) \in R_{-(\gamma+1)}$. 

According to Gagnon and Hayashi (2023), for any outlier ($y_o$) and fixed $(\boldsymbol\beta, \sigma)$,
\begin{align}\label{eqn:limit_PDF}
 \lim_{\omega \rightarrow \infty} \frac{(1 / \sigma) g((y_o) / \sigma)}{g(y_o)} = \sigma^\gamma.
\end{align}
Using this property, we have 
 \begin{eqnarray}
 \label{Eq3.12}
  \lim_{\omega \rightarrow \infty} \pi(\sigma| \textbf{X},\textbf{y})&=&\frac{\sigma^{\gamma}\pi(\sigma|X^{(n-1)},\textbf{y}^{(n-1)})}{\lim_{\omega \rightarrow \infty}\int_{0}^{\infty} \frac{1}{\sigma}g\left(\frac{y_n}{\sigma}\right)/g\left(y_n\right)\pi(\sigma|\textbf{X}^{(n-1)},\textbf{y}^{(n-1)}) d\sigma}.   
\end{eqnarray}
From (\ref{Eq3.8}), we obtain
 \begin{eqnarray}
\pi(\sigma|\textbf{X}^{(n-1)},\textbf{y}^{(n-1)}) \propto \sigma^{-(n-p)} l(\sigma).
 \end{eqnarray}
By applying Potter's theorem, the condition for the inside of the dominator for some $C$ and any $\delta$ can be expressed as follows.
\begin{eqnarray}
\frac{1}{\sigma}g\left(\frac{y_n}{\sigma}\right)/g\left(y_n\right) \leq C \cdot \max\{ \sigma^{\gamma+\delta}, \sigma^{\gamma-\delta} \}.
 \end{eqnarray}
According to Andraid and O'Hagan (2006),
if $l \in R_{0}$ is measurable and $\alpha<-1$, for $x=(0,\infty)$,
\begin{eqnarray}
\int^{\infty}_0 x^{\alpha} l(x) dx< \infty.
\end{eqnarray}
Thus, for the dominator of (\ref{Eq3.12}) to exist, the condition (1b), $\gamma<(n-p-1)$, must hold. 
\qed
\\

Now, we extend the model with multiple outliers. Let L and K be the datasets of outliers and non-outliers, respectively. The numbers of outlying and non-outlying data points are given as $l \in \mathbb{N}$ and $k \in \mathbb{N}$, respectively. Thus, $n=l+k.$\\

 \noindent \\
\bf{Theorem 2 (Robustness of $l$ outliers among $n$ observations).}
\it{Consider data \rm$\textbf{y}=(\it y_1, \ldots,y_n)$ following model \rm (\ref{Eq333})\it,  and $l$ outliers out of $n$ observations. If Lemmas 1 and 2, Corollary 1 hold for each outlier, and the following conditions hold:

\rm(2a)  $(\gamma+1) \cdot l < \{ (\gamma+1) \cdot k \}$,

\rm(2b)   $l \cdot \gamma  <  ( k-p )$;   

\it

, the posterior distribution partially ignores the outliers.
\begin{equation}
\label{Eq3.13}
\pi(\beta,\sigma|\mathbf{X}, \mathbf{y}) \propto \sigma^{l \cdot \gamma}\pi(\beta, \sigma | \bf {X^{\rm (K)}}, \bf{{y}^{\rm (K)}}) \hspace{1cm} \textrm{as} \hspace{0.3cm} \it \omega \rightarrow \infty,
\end{equation}
where the superscript \rm (K) \it indicates the dataset of non-outliers.  \\ 
 
\noindent
Proof.
\rm
When a scale parameter $\sigma$ is given, the posterior distribution of $\beta$ in the model is as follows:
\begin{eqnarray}
\label{Eq3.14}
\pi(\beta|\sigma, \bf {X^{\rm (K)}}, \bf{y}^{\rm (K)})  \in \textit R_{- \textit{k} \cdot (\gamma+\textrm{1})}.
\end{eqnarray}

Thus, from the condition $l<k$, even when all outliers are sufficiently close and move in the same direction, where outliers create a lighter distribution, the condition (2a) is satisfied.

The marginal posterior distribution of $\sigma$ given information $\bf {X^{\rm (K)}}$ and $\mathbf{y}^{\rm (K)}$ becomes
\begin{eqnarray}
\label{Eq3.15}
\pi(\sigma| \bf {X^{\rm (K)}}, \mathbf{y}^{\rm (K)})\in \textit R_{- (\textit{k-p}+\textrm{1})}, 
 \end{eqnarray}

whereas that of $\sigma$ for all observations is 
 \begin{eqnarray}
 \label{Eq3.16}
 \pi(\sigma| \textbf{X}, \textbf{y})=\frac{\{\Pi_{y_i \in L}f(y_{i}|\sigma, \textbf{X}^{\rm(K)},\textbf{y}^{\rm(K)}) \} \pi(\sigma|X^{\rm(K)},\textbf{y}^{\rm(K)})}{\int_{0}^{\infty} \{\Pi_{y_i \in L}f(y_{i}|\sigma, \textbf{X}^{\rm(K)},\textbf{y}^{\rm(K)}) \}\pi(\sigma| \textbf{X}^{\rm(K)},\textbf{y}^{\rm(K)}) d\sigma}.
\end{eqnarray}
We have 
 \begin{eqnarray}
 \label{Eq3.17}
  \lim_{\omega \rightarrow \infty} \pi(\sigma| \textbf{X},\textbf{y})&=&\frac{\sigma^{l \cdot \gamma}\pi(\sigma|X^{\rm(K)},\textbf{y}^{\rm(K)})}{\lim_{\omega \rightarrow \infty}\int_{0}^{\infty}\{\Pi_{y_i \in L}\{ \frac{1}{\sigma}g\left(\frac{y_i}{\sigma}\right)/g\left(y_i\right)\}\pi(\sigma|\textbf{X}^{\rm(K)},\textbf{y}^{\rm(K)}) d\sigma}.   
\end{eqnarray}

Thus, for the dominator of (\ref{Eq3.17}) to exist, the condition (2b), $ l \cdot \gamma<(k-p)$, must hold. 
\qed
\\

If $\gamma \geq 1$, the sufficient conditions becomes only $ l \cdot \gamma<(k-p)$, as well as in Theorem 1.\\

\noindent
\it Remark. \rm When $n$ is sufficiently greater than $p$, the breakdown point of the Student-$t$ linear regression modeling with $t$ distribution with 3-DoFs is robust when the ratio of outliers is less than approximately $25\%$.

\rm{
\subsection{Example}
 
We consider the following case of a simple linear regression model with a single outlier:  
 \begin{eqnarray}
 \label{Eq4.1}
y_i =\beta_0+\beta_1x_i + u_i       \hspace{1cm}    (i=1, \ldots ,n).
\end{eqnarray}
We use a simple regression to visualize our conditions for robustness, which we provided in Sections 2 and 3. This is because when we apply our limitation on multiple regression, it is difficult to illustrate the visualization.

\color{black}
From Lemma 1, we obtain the following condition:
\begin{eqnarray}
 \label{Eq4.2}
h_{nj}& =&\frac{1}{n}+\frac{(x_{out}-\bar{x})(x_{j}-\bar{x})}{\sum^{n}_{i=1}(x_{i}-\bar{x})^2}\nonumber \\
& =&\frac{1}{n}+\frac{\frac{n-1}{n}(x_{out}-\bar{x}^*)(x_{j}-\bar{x}^*)-\frac{n-1}{n^2}(x_{out}-\bar{x}^*)^2  }{\sum^{n-1}_{i=1}(x_{i}-\bar{x}^*)^2+\frac{n-1}{n}(x_{out}-\bar{x}^*)^2} \nonumber \\
& =&\frac{1}{n}+\frac{(x_{out}-\bar{x}^*)(x_{j}-\bar{x}^*)-\frac{1}{n}(x_{out}-\bar{x}^*)^2  }{\frac{n}{n-1}\sum^{n-1}_{i=1}(x_{i}-\bar{x}^*)^2+(x_{out}-\bar{x}^*)^2} \nonumber \\
& >&0,
\end{eqnarray}
where $\bar{x}^*=\frac{1}{n-1}\sum^{n-1}_{i=1}x_{i}$. 

By arranging condition (\ref{Eq4.2}), we obtain the range as 
\begin{eqnarray}
 \label{Eq4.3}
-(x_{out}-\bar{x}^*)(x_{j}-\bar{x}^*)<  \frac{1}{n-1} \cdot \sum^{n-1}_{i=1}(x_{i}-\bar{x}^*)^2 .
\end{eqnarray}

Thus, when $(x_{out}-\bar{x})>0$, Lemma 1 is satisfied in the range:
\begin{eqnarray}
\label{Eq4.4}
-(x_{j}-\bar{x}^*)<  \frac{(x_{out}-\bar{x}^*)}{\frac{1}{n-1} \cdot \sum^{n-1}_{i=1}(x_{i}-\bar{x}^*)^2 }.
\end{eqnarray}

 The above condition shows that all non-outliers need to be closer than the standardized distance, standardized by the standard deviation of the non-outliers, between the outlier and the average of the non-outliers.

\color{black}
 From Lemma 2, we obtain the following condition:
\begin{eqnarray}
 \label{Eq4.5}
h^{out}_{nn}& =&\frac{1}{n}+\frac{(x_{out}-\bar{x})^2}{\sum^{n}_{i=1}(x_{i}-\bar{x})^2}\nonumber \\
& =&\frac{1}{n}+\frac{\left(\frac{n-1}{n}\right)^2(x_{out}-\bar{x}^*)^2  }{\sum^{n-1}_{i=1}(x_{i}-\bar{x}^*)^2+\frac{n-1}{n}(x_{out}-\bar{x}^*)^2} \nonumber \\
& =&\frac{1}{n}+\frac{\frac{n-1}{n}(x_{out}-\bar{x}^*)^2  }{\frac{n}{n-1}\sum^{n-1}_{i=1}(x_{i}-\bar{x}^*)^2+(x_{out}-\bar{x}^*)^2} \nonumber \\
& <&\frac{1}{2}.
\end{eqnarray}
By rearranging the above condition, we obtain the range as 
\begin{eqnarray}
 \label{Eq4.6}
(x_{out}-\bar{x})^2<  \frac{n-2}{n-1} \cdot \sum^{n-1}_{i=1}(x_{i}-\bar{x}^*)^2 .
\end{eqnarray}

This condition indicates that the distance between the outlier and the average of all observations needs to be less than $(n-2)$ times the standard deviation of the non-outliers.
\color{black}

Thus, when $(x_{out}-\bar{x})>0$, Lemma 2 can be satisfied in the range
\begin{eqnarray}
\label{Eq4.7}
 \bar{x}^*- \left [ (n-2) \cdot \frac{\sum^{n-1}_{i=1}(x_{i}-\bar{x}^*)^2 }{n-1} \right ]^\frac{1}{2}<x_{out} <\bar{x}^*+\left [ (n-2) \cdot \frac{\sum^{n-1}_{i=1}(x_{i}-\bar{x}^*)^2 }{n-1} \right ]^\frac{1}{2}
\end{eqnarray}

These results highlight that the robust range is wider, as the number of non-outliers is greater. In addition, when the independent variable of the outlier is located far from other data, no conflict of information exists, irrespective of the value of $y$.

\rm{
\subsubsection{Simulation results}

This subsection investigates the robustness performance in relation to the value of the outlier in $X$ for the Student-$t$ linear regression model. We observe the impact on the posterior mean of $\beta_1$ for different values of covariate $X$.\\

\color{black}
\noindent
\bf{The simulated observations} \rm

To satisfy Corollary 1, we use the following two datasets for the covariate $X$ of the non-outliers.\\

Simulation 1 \quad \quad \quad $x_{/out}=[ -2, -1, 0, 1, 2]$.\\

Simulation 2 \quad \quad \quad $x_{/out}=[ -2, -1, 0, 1, 2,  -2, -1, 0, 1, 2]$.\\ 

We generate the non-outliers to be sufficiently close to the regression line, we set relatively small error term as follows:
\begin{eqnarray}
y_i=3+2 x_i +u_i   \hspace{0.3cm}(i=1,\ldots,n),
\end{eqnarray}
where $u_i \stackrel{D}{\sim} N(0,1)$.

For the outlier, we move the outlier in the $x$-direction, from $-50$ to 50. We set $y_{out}$=$-10$ to locate far enough from the non-outliers. Figure 3 illustrates the data used for these simulations.\\

\color{black}

\begin{figure}[h]
\begin{center}
\includegraphics[ width=60mm]{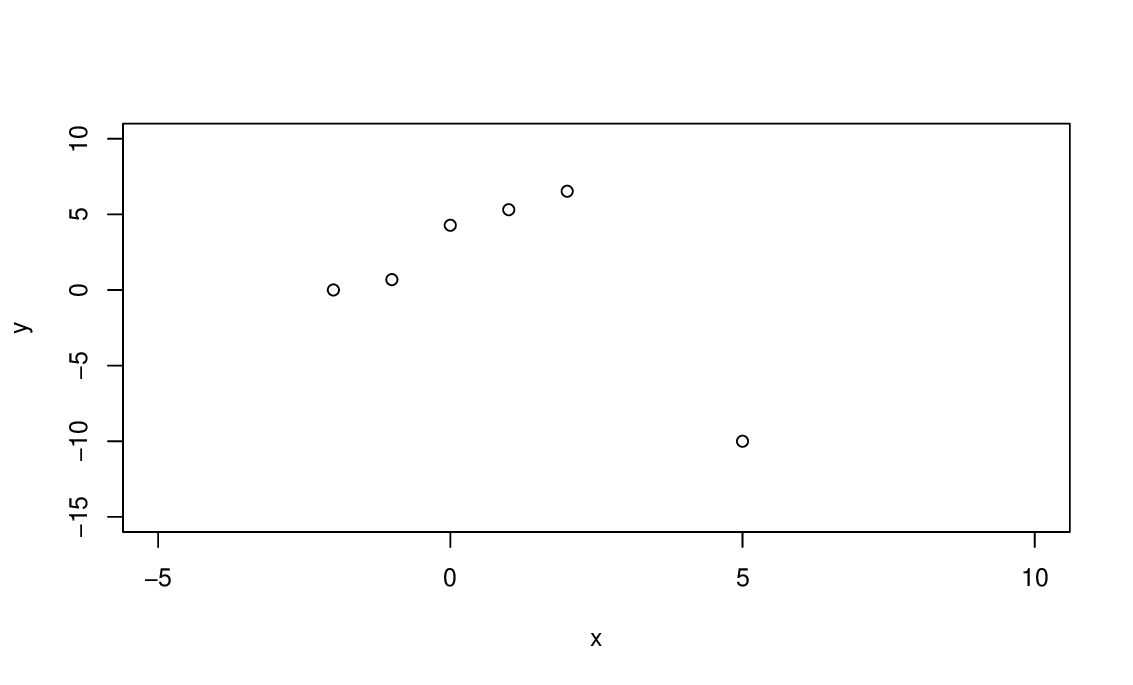}
\includegraphics[ width=60mm]{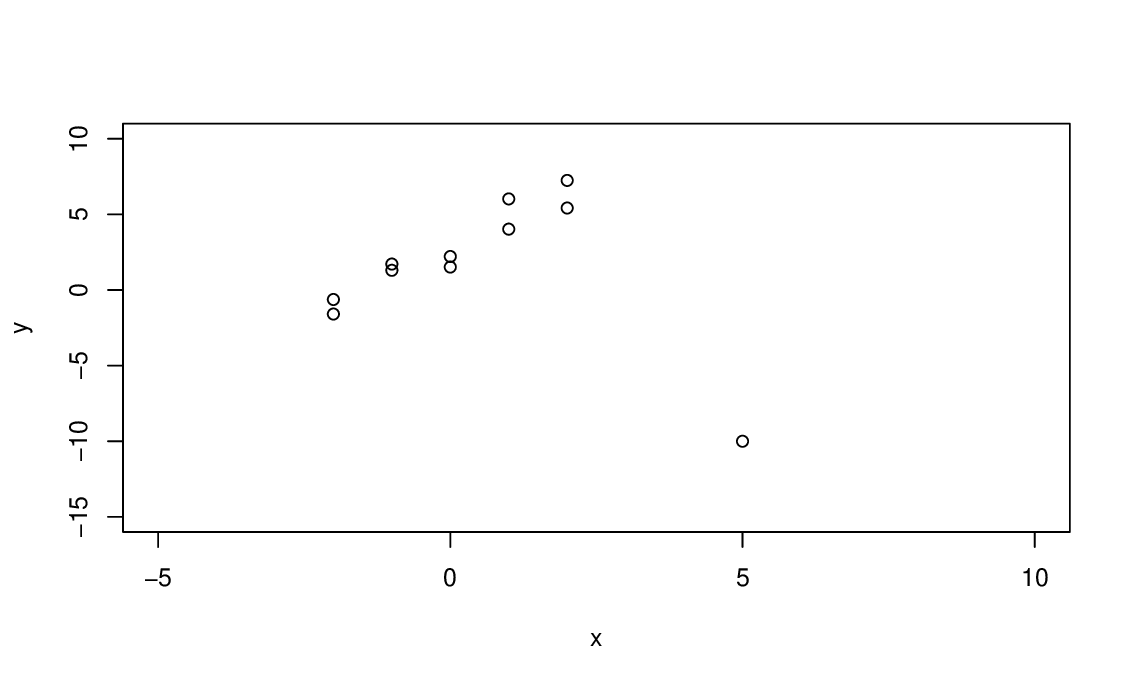}
\vspace{0cm}
\caption{Scatterplots of simulated data. }
\end{center}
\end{figure}

\noindent
\bf{Estimation Model} \rm

We employ the independent Jeffreys priors; the priors of $\beta_0$ and $\beta_1$ have uniform distributions, and the prior distribution of $\sigma$ is $1/\sigma$. When the sample size is relatively small to the DoFs of the $t$-distributed errors, the convergence is slow. Thus, we utilize 3-DoFs, $u_{i} \sim t_{(3)}(0,\sigma)$ for the error term of the likelihood. From Theorem 1, the sufficient condition for a $t$-distribution with $\gamma$ DoFs is $\gamma <\{ n-p-1\}$. Thus, in this model, the condition becomes $3<n-4$.\\
 
\noindent
\bf{Simulation results} \rm

Figure 4 illustrates the results of the numerical evaluation of the posterior mean of parameter $\beta_1$. The upper right panel illustrates the result for $n=6$, satisfying the sufficient condition, and the lower right panel presents the result for $n=11$, satisfying the condition. The results indicate that the Student-$t$ linear regression model is robust within the controllable range defined in Lemma 2, represented as the vertical dotted lines. 

\begin{figure}[h]
\begin{center}
\includegraphics[ width=60mm]{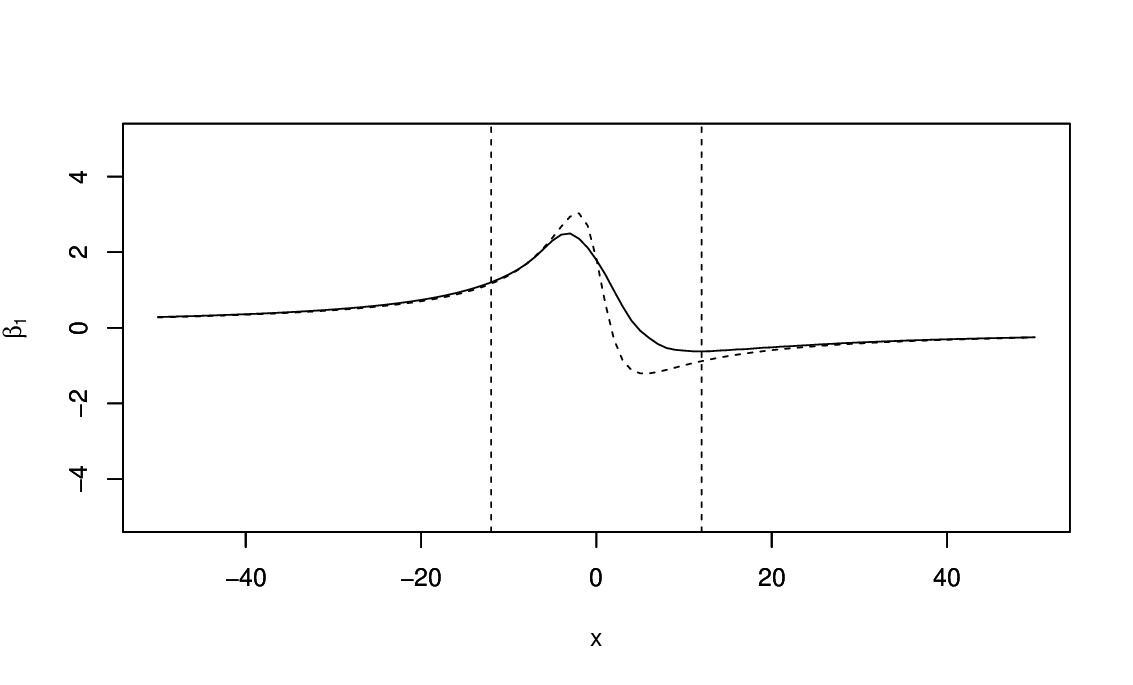}
\includegraphics[ width=60mm]{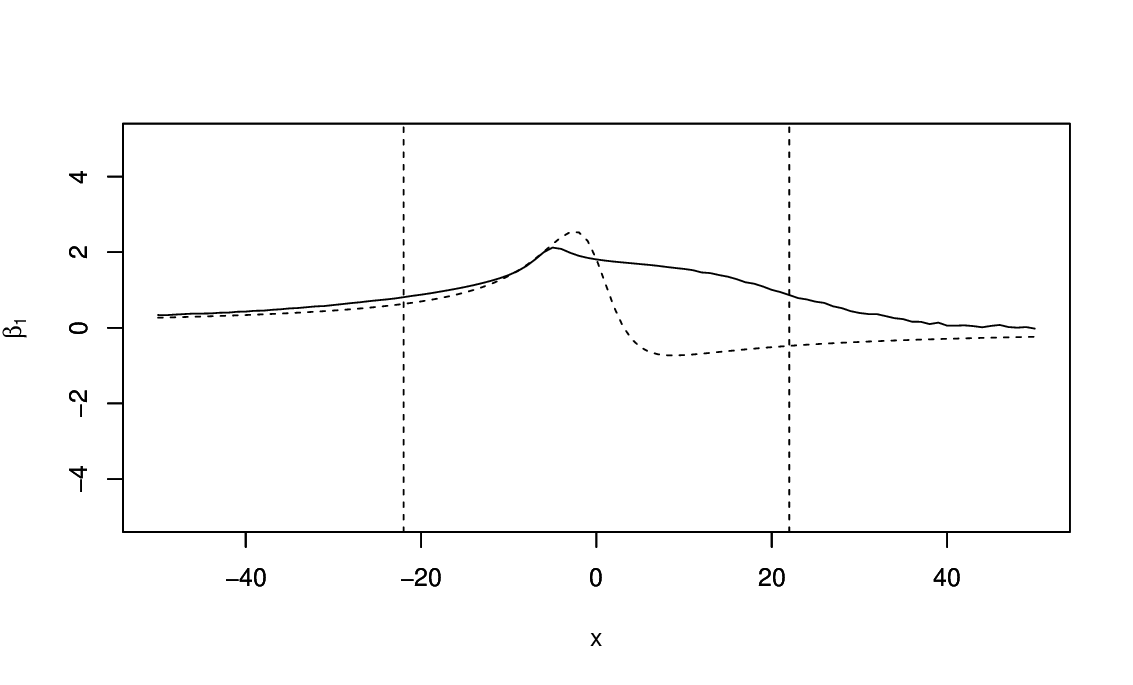}
\vspace{0cm}
\caption{Posterior mean of the slope $\beta_1$. The upper panel shows the result for $n=6$, whereas the lower panel shows the result for $n=11$. The straight line depicts the result of the Student-$t$ linear regression model, whereas the dashed line depicts that of the linear regression model with normally distributed error terms. The vertical dotted lines show the range defined in Lemma 2.}
\end{center}
\end{figure}

 \section{Concluding remarks}

This study extended Andrade and O'Hagan's (2011) condition 
for addressing the outlier problem of the Student-$t$ linear regression model. The condition is efficient when conflicting information exists between outliers and non-outliers. However, in a linear regression model, an outlier does not conflict with non-outliers when it is located far from them in the $x$-direction. Thus, we first clarified the range of the presence of conflicting information in a linear regression model using Hat matrix with the elements of $h_{ij}$ as $h_{nj}>0$, $h_{nn}<\frac{1}{2}$ and $\Sigma_{d \neq j}^{n-1}(h_{nd})^2 \geq \frac{1}{8}$ for the model with the $n$th observation being an outlier. The results show that the Student-$t$ linear regression model is only robust against outliers if certain limitations on $X$ are met.

Thereafter, we established the sufficient conditions for robustness of the Student-$t$ linear regression model with multiple outliers in the range, as $l<k$ and $l \cdot \gamma  <  ( k-p )$ with $l$ outliers and $k$ non-outliers. \color{black} Future studies should investigate extending this study to a model with unknown DoFs for the $t$-distribution.\\

\noindent
\bf{ Acknowledgments}

\rm This study was supported by the Women Researchers Support Office of Osaka Metropolitan University Grant. Also, the author thanks  Prof. Yoshihiko Konno as well as two anonymous referees for their helpful suggestions that led to an improved manuscript.

\newpage
\noindent
\textbf{Appendix A: Regularly varying functions}
\renewcommand{\theequation}{A.\arabic{equation} }
\setcounter{equation}{0}

The tail behavior can be presented by the index of a regularly varying function. Index $\rho$ is defined as follows.

A positive measurable function $f(x)$ is regularly varying at $\infty$ with index $\rho \in R$ for an arbitrary positive $t$.
\begin{eqnarray}
\lim_{x \rightarrow \infty} \frac{f(t x)}{f(x)}=t^{\rho}.
\end{eqnarray}

We present it as $f(x) \in R_\rho$ in this study, and $l(x) \in R_0$ is known as ``slowly varying.'' The regularly varying function can be presented as $f(x)=t^\rho l(x)$.

Using the property
\begin{eqnarray}
\lim_{x \rightarrow \infty} \frac{\log (f(x))}{\log x}=\rho,
\end{eqnarray}

we obtain the index for $t$ distribution with the DoFs, $\gamma$, as

\begin{eqnarray}
\lim_{x \to \infty}\frac{\log(p(x;\gamma,\mu,\sigma^2))}{\log x} 
&=&\lim_{x \rightarrow \infty} \frac{(A-\frac{\gamma+1}{2}\log(\{1+\frac{1}{\gamma}(\frac{x-\mu}{\sigma})^2\}))}{ \log x}\\ \nonumber
&=&-(\gamma+1),
\end{eqnarray}

where $A=\log\left(\frac{\Gamma(\frac{\gamma+1}{2})}{\Gamma(\frac{\gamma}{2}) \pi^{1/2}\gamma^{1/2}\sigma}\right)$.\\

 Some properties of the regularly varying function used in this study are highlighted in Bingham et al. (1987) and Resnick (2007).\\

\end{document}